# Direct observation of quantum anomalous vortex in Fe(Se,Te)

Y. S. Lin[1#], S. Y. Wang[1#], X. Zhang[2], Y. Feng[1], Y. P. Pan[1], H. Ru[1], J. J. Zhu[1], B. K. Xiang[1], K. Liu[1], C. L. Zheng[1], L. Y. Wei[2], M. X. Wang[2,3], Z. K. Liu[2,3], L. Chen[4], K. Jiang[5], Y. F. Guo[2], Ziqiang Wang[6], Y. H. Wang[1,7*]

*1. State Key Laboratory of Surface Physics and Department of Physics, Fudan University, Shanghai 200433, China*
*2. School of Physical Science and Technology, ShanghaiTech University, Shanghai 201210, China*
*3. ShanghaiTech Laboratory for Topological Physics, Shanghai 201210, China*
*4. Shanghai Institute of Microsystem and Information Technology, Shanghai 200050, China*
*5. Institute of Physics, Chinese Academy of Sciences, Beijing 100190, China*
*6. Department of Physics, Boston College, Chestnut Hill, MA 02138, USA*
*7. Shanghai Research Center for Quantum Sciences, Shanghai 201315, China*

\# These authors contributed equally.
\* To whom correspondence and requests for materials should be addressed. Email: wangyhv@fudan.edu.cn

**Vortices are topological defects of type-II superconductors in an external magnetic field. In a similar fashion to a quantum anomalous Hall insulator, quantum anomalous vortex (QAV) spontaneously nucleates due to orbital-and-spin exchange interaction between vortex core states and magnetic impurity moment, breaking time-reversal symmetry (TRS) of the vortex without an external field. Here, we used scanning superconducting quantum interference device microscopy (sSQUID) to search for its signatures in iron-chalcogenide superconductor Fe(Se,Te). Under zero magnetic field, we found a stochastic distribution of isolated anomalous vortices and antivortices with flux quanta $\Phi_0$. By applying a small local magnetic field under the coil of the nano-SQUID device, we observed hysteretic flipping of the vortices reminiscent of the switching of ferromagnetic domains, suggesting locally broken-TRS. We further showed vectorial rotation of a flux line linking a paired vortex-**

**antivortex with the local field. These unique properties of the anomalous vortices satisfied the defining criteria of QAV. Our observation suggests a quantum vortex phase with spontaneous broken-TRS in a high-temperature superconductor.**

Vortices are singularities in the phase winding of the complex order parameter of a superconductor. Abrikosov vortices with quantized fluxoid nucleate when an external field $H$ exceeds the lower critical field $H_{c1}$ or when cooling through the critical temperature $T_c$ in a finite $H$ [1]. Spontaneously generated vortices without any applied magnetic field, or anomalous vortices, requires breaking time-reversal symmetry (TRS) inside the superconducting ground state [2]. Breaking TRS in a superconductor is a non-trivial effect which leads to exotic excitations [3–5]. Certain heavy fermion superconductors have spin-triplet pairing which can break TRS [6–8]. Spontaneous boundary current, yet to be found, is a direct manifestation of the underlying TRS breaking. Spontaneous vortices thus signify a similar origin as they are associated with supercurrent around interior boundaries of the topological defects.

Superconductors with strong spin-orbit interaction and exchange coupling offer another route to break TRS. The mechanism is in analogy with a quantum anomalous Hall (QAH) state in a magnetically-doped thin-film topological insulator [9,10]. The QAH originates from combining the spin-orbit interaction of the topological insulator with the exchange interaction of the magnetic impurities. Spontaneously broken-TRS in the bulk leads to unique emergent properties of a QAH insulator that are directly observable: magnetic domains of opposite polarization appearing after cooling under zero-field [11], persistent chiral boundary current and quantized Hall conductance whose signs are determined by the sign of the magnetization [9,10], and hysteretic magnetization under external field sweeps. All these defining properties of QAH can be used to distinguish spontaneous vortices in superconductors generated quantum mechanically from those that occur thermally. The quantum regime is reached by a large superconducting gap relative to the Fermi energy which maintains well-separated vortex core states at finite temperature. The magnetic impurity induces circulating current which modulates the phase of superconducting order parameter. Strong spin and orbital exchange interactions make it energetically favorable to nucleate a vortex around the impurity moment compared to a vortex-free state.

The iron-chalcogenide superconductor FeSe$_{0.5}$Te$_{0.5}$ (FST) [12–20] is such an unconventional superconductor with these key ingredients. For a superconducting gap of 1.5 meV and Fermi energy of 4.52 meV on the hole-band around Γ [21], its normal vortex core-states are separated by 0.76 meV [22], sufficiently large for its critical temperature of 16 K [23]. It has strong spin-orbit coupling and a topological bulk band in its normal state [21,24–26]. Possible gap-opening were observed at the surface Dirac point by photoemission spectroscopy, suggesting broken-TRS at the surface [27]. We note, however, the Dirac surface states are not essential for QAV [2], although the Dirac cone adds the potential to generate Majorana zero modes. Therefore, zero-energy modes observed in cores of Abrikosov vortices in FST [19,22,25,26,28,29] are irrelevant for QAV. Instead, the observation of QAV, which is a flux quantization and vorticity phenomenon, relies on direct measurements of quantized magnetic flux in a zero-field environment and its hysteretic response under a field sweep. All these conditions demand highly sensitive scanning magnetic probes.

Scanning superconducting quantum interference device (sSQUID) microscopy [30–33] is a very sensitive and direct flux-imaging technique. We have installed magnetic shielding around the sample and the SQUID to minimize the magnitude of the external magnetic field (such as the geo magnetic field). For any residual field after the shielding, which is on the order of mG, we used a home-made superconducting coil around the sample to compensate. These procedures allowed us to accurately determine the zero magnetic field condition [23]. The field coil (Fig.1a) on the nano-SQUID [34] can apply a local magnetic field to manipulate a vortex/antivortex without affecting the global vortex structure or suppressing superconductivity.

There were trapped vortices in an FeSe$_{0.5}$Te$_{0.5}$ flake sample when the applied flux $\Phi_a$ was relatively large. We followed the exact same field cooling and measurement procedure as we did on a conventional superconductor with similar diamagnetic strength which also calibrated the external field and $\Phi_a$ [23]. Previous studies showed that the chemical composition and the superconductivity of FST were phase separated into domains hundreds of microns in size [35]. We avoided such inhomogeneity by choosing suitable pieces from the small flake samples after exfoliation. Since the diamagnetic susceptibility was directly proportional to superfluid density, the uniformity in susceptometry (Fig. 1a) showed superconducting homogeneity of our sample. Our extensive sample characterizations at different length scales also suggested uniform superconductivity up to the sample size [23]. We cooled the FST sample from 18 K to 1.5 K through its $T_c$ ~ 14 K at a cooling rate of 50 mK/s to prevent any thermally-excited vortex-antivortex pairs from freezing [36]. We applied a fixed $\Phi_a$ during cooling and scanning microscopy and repeated the process for various $\Phi_a$'s. In the regime of $|\Phi_a| > \Phi_0$, where $\Phi_0 = h/2e$ ($h$ is the Planck's constant and $e$ is the electron charge) is the superconducting flux quantum, we obtained the magnetometry images showing trapped vortex throughout the sample (Figs. 1b-e). These vortices exhibited the same vorticity determined by the sign of $\Phi_a$. This situation seemed similar to that of a conventional superconductor (Figs. S6) except that the number of observed vortices was clearly larger than $|\Phi_a|/\Phi_0$ in FST. The average value and the spread of the width of the vortices were very similar to those of vortices on the Nb film (Fig. S11). This fact was a further testament of the superconducting homogeneity of the FST samples and suggested that large fields predominantly induced Abrikosov vortices.

Surprising vortex patterns appeared when $|\Phi_a| < \Phi_0$, the low-field regime where no vortex was expected or observed on a conventional superconductor (Fig. S6). We performed 27 times of field-cooling cycles in this regime on this particular FST sample [23]. The magnetometry images were drastically different even if $\Phi_a$ was changed slightly and we present here some representative ones (Figs. 2a-f). Vortices of both vorticity showed up simultaneously within the same image even for finite $\Phi_a$ (Figs. 2a-c). At calibrated zero field, the images were different each time: in some cases, no vortex was observed (Fig. 2d), while in others vortices and antivortices appeared at random locations (Figs. 2e and f). The fact that the zero-vortex state did occur ruled out thermally-induced vortex-antivortex, which always generated the same density of pairs at a fixed quenching rate [36]. Regardless of vorticity, isolated vortices showed total flux of $\Phi_0$ (Figs. 2a, b and f). We have observed similar effect in all the FST samples we fabricated with similar composition [23].

The FST samples we used contained a dilute amount of interstitial Fe (<1%) to act as impurity magnetic moments. At such low concentration, the impurity moments did not form long-range magnetic order [12-15]. Nevertheless, chemical or crystalline inhomogeneities may cause flux-trapping in even well-shielded low field environment. For these reasons, we carried out similar zero-field cooling and measurements in the same setup in different types of control samples: Fe(2 nm)/Nb(80 nm), FeSe without the Te alloying and $Fe_{1+y}(Se,Te)$ with a much higher concentration (nominally ~10%) of interstitial Fe impurity [23]. Since these control samples had either nominally similar or higher impurity density than FST, they further ruled out conventional pinning due to sample inhomogeneity [37,38] or instrumental artefact. The absence of vortex in FST above $T_c$ ruled out magnetic clustering. These control experiments suggest that the stochastic occurrence of spontaneous vortices and antivortices with quantized flux under zero magnetic field was an intrinsic effect of FST with low interstitial Fe.

The number of vortices and antivortices from all these cooling cycles gave us statistical insights of the peculiar random vortex patterns. When $|\Phi_a| < \Phi_0$, the number of vortices/antivortices in FST fluctuated towards both sides in similar amplitudes against a flat baseline from the Nb control sample (Fig. 2g). However, as $|\Phi_a|$ got bigger, there was a positive correlation between $\Phi_a$ and the number of vortices with the same sign as $\Phi_a$. Although the probability of observing antivortices quickly diminished when $|\Phi_a| \gtrsim \Phi_0$, the number of observed vortices still fluctuated. The total number of observed vortices was much higher than $|\Phi_a|/\Phi_0$ (Fig. 2g, black dashed line), whereas the number of Abrikosov vortices in Nb was lower. Such behavior was consistent with a positive bulk magnetization below $T_c$ under field cooling of FST with magnetic impurities [20]. The generation of more flux than what was applied suggested ferromagnetic exchange interaction from the magnetic impurity moment, such as that of the excess Fe ions.

Having established the existence of anomalous vortex and its connection with magnetic impurity, we study the flux profiles of spontaneous vortex-antivortex duos to understand the origin of randomness. Duos which located close to each other (Figs. 2a, c and f, arrows) showed quite different contrast and shape from the isolated ones (Figs. 2a, b and f, dashed circles). The line-cuts through the duos showed that the peak-to-peak flux reduced with decreasing distances (Fig.

2h). The farthest duo (v1) has about 0.9 $\Phi_0$ per vortex and looked like two monopoles with opposite signs (Fig. 2a). The closest duo (v3) amounts to only 0.15 $\Phi_0$ per vortex and appeared as a magnetic dipole (Fig. 2h). Out of the 27 images, there were 6 discernable cases of v3 [23]. The variation in the shape of the duo could be understood when considering the magnetic energy of the flux line. The decay length we observed of a typical isolated vortex was around 10 μm (Fig. 2b), consistent with the expected Pearl length $\Lambda = \frac{2\lambda^2}{d}$, where $\lambda \sim 500$ nm was the London penetration depth of bulk FST and $d \sim 100$ nm [23] was the thickness of our sample. When the pair separation was larger than $\Lambda$, there was little magnetic interaction between the vortex and antivortex. The flux lines through them were mostly normal at the surface, bending to connect far away from the sample. The quantized flux captured by sSQUID close to the surface was thus hardly affected by the other and the duo assembled into a double-monopole pattern. As the separation got smaller, it costed more energy for the flux lines to bend around a 'U-turn'. As a result, the flux lines tilted toward each other slightly at the surface to reduce the bend, which led to reduced flux along the surface normal (v2). When the separation was further reduced (v3), the flux lines tilted heavily toward each other to minimize the magnetic energy. This diminished the flux along the surface normal and the flux of the vortex appeared $\ll \Phi_0$. Since the average distance between interstitial Fe ions [15], through which vortex/antivortex resided, was much smaller than $\Lambda$, most of the duos canceled each other out on the mesoscopic scale. Those that were detected happened because the two properly separated vortex and antivortex were not paired up by others that were much closer. This process of pair-making from statistically large amount of impurity centers led to the stochastic nature of the anomalous vortex patterns.

In order to study the switching of vortex-antivortex pair under magnetic field, we focused on a different FST sample (Fig. 3). After zero-field cooling, this sample had a higher probability of forming the dipole-like vortex-antivortex structure in its lower section (Fig. 3a), likely due to its particular shape. To tune the moments of the vortices without affecting other part of the sample, we applied a local magnetic field $H_F$ by passing a current through the field coil on our nano-SQUID (Fig. 3a). Before each scan, we moved the field coil over to the middle of the duo and applied a particular $H_F$ as labelled above the images. Then, we turned $H_F$ to 0 before acquiring the magnetometry image. Each image (Figs. 3a-j) was obtained in this manner following the sequence of 'down-sweep' and then 'up-sweep', completing a loop. Starting from $H_F = 57.6$ G (Fig. 3a), the vortex with a positive flux appeared on the left side of the antivortex enclosing a negative flux. As the applied $H_F$ decreased, the contrast of the dipole reduced at $H_F = 18$ G (Fig. 3b) but kept the same sign even after a reversed $H_F = -21.6$ G was applied (Fig. 3c). When $H_F = -28.8$ G, the dipole pattern reversed its sign with a much-reduced contrast (Fig. 3d). The contrast started to increase when $H_F = -43.2$ G (Figs. 3e) and peaked out around $H_F = -57.6$ G (Fig. 3f), reaching a reversed pattern of the initial state (Fig. 3a). The up-sweep images (Figs. 3f-j) were just the opposite of the down-sweep ones, where reversal appeared at $H_F = 36$ G (Fig. 3i). The pattern we obtained after cycling back to $H_F = 57.6$ G (not shown) was exactly the same as the original one. It was clear from the above sequence that the vorticities of the vortex and antivortex under a local field was history-dependent.

The hysteresis in the switching was better visualized in a flux-field diagram (Fig. 3k). The trajectories followed by the vortex and the antivortex followed opposite winding directions with respect to the field sweep. The flux of both loops reversed signs after the field switched directions. By combining the line cuts through the dipole taken from magnetometry images at various $H_F$ of the down-sweep (Fig. 3l) and that of the up-sweep (Fig. 3m), we found that the vortex and antivortex switched their signs concurrently. Such synchronous inversion of their vorticity suggested that they were paired by threading a common flux tube through both impurity moments. Conventional vortices with pinning also exhibit hysteresis in the magnetization-field curve. However, the magnetization typically reverses sign before the field ramps down to zero [1,39–41]. Instead, our anomalous vortex loops were reminiscent of a magnetization-field hysteresis loop of a ferromagnet. (Note that the reversal fields were not as sharp as a typical coercive field of a hard ferromagnet largely because the local field $H_F$ had to be removed during scanning imaging to avoid scrambling the structure.) The ferromagnet-like hysteresis loop strongly suggested spontaneous TRS-breaking. The anomalous vortex and antivortex we observed satisfied all the signatures of QAV.

The magnitude and polarities of the pair revealed vectorial rotation of magnetic impurities moments under the application of $H_F$. The total flux of the vortex (antivortex) was 1.0 $\Phi_0$(−0.6 $\Phi_0$) when the contrast reached its maximum at $H_F = 57.6$ G (Fig. 3a), whereas that at the reversal field $H_F = -28.8$ G (Fig. 3d) was 0.1 $\Phi_0$(−0.2 $\Phi_0$). These values corresponded to cases v2 and v3 pairs respectively and suggested that they had an in-plane component. The fact that the total flux of the pair did not sum to zero suggested that the moments in the vortex cores were not antiparallel. The tilting from the surface normal into the plane was most obvious when the pair has the least out-of-plane component, which happened at the reversals. The images we obtained at $H_F = -28.8$ G during the down-sweep (Fig. 4a) and $H_F = 36$ G during the up-sweep (Fig. 4b) showed qualitatively similar patterns. They were both consisted of two dipoles aligned off-axis and their polarities followed the same 'minus-plus, minus-plus' order clockwise (Fig. 4c). The characteristic $H_F$ is an order of magnitude smaller than the $H_{C1}$ of FST at 3 K [23] and thus not capable of generating or annihilating an Abrikosov vortex. The Meissner current induced by such a small local magnetic field exerts a Lorentz force on an existing Abrikosov vortex and causes some lateral displacement [42]. However, the dipolar pattern and the hysteretic switching were both inconsistent with a lateral displacement of the original structure.

Remarkably, each dipolar pattern agrees well with the theoretical calculations of a single QAV nucleated at a magnetic impurity carrying a local moment canted away from the normal of the surface [43]. To simulate the out-of-plane field in the above configuration, we combined the calculated current distributions of isolated QAV and antivortex with in-plane moments [23] and computed the out-of-plane field measured by sSQUID using Biot-Savart law (Fig. 4d). For such a crude model without explicitly including the vortex-antivortex interaction, it showed qualitative agreement with the measured patterns (Figs. 4a and b). The agreement suggested that the Meissner current induced by $H_F$ was driving a vectorial rotation of the coupled spin-flux line (Fig. 4e). This sequence of vortex-antivortex pair rotation by small $H_F$ is a manifestation of the interaction between the impurity spin and the orbital motion of the vortex core states, which is the origin of QAV's quantum nature.

The emergent behavior of the QAV was a local effect in the low excess Fe impurity (<1%) samples we measured. In this regime, the vortex cores were still well-separated such that formation of vortex-antivortex was more favorable than parallel alignments of the moments due to the lower free energy of the pair. Lowering the Fe impurity would reduce the probability of finding the QAV. On the other hand, increasing the Fe impurity density would decrease their average distance to be much smaller than the coherence length so that the vortex cores would start to overlap heavily. They might develop long-range ferromagnetic order by an RKKY-like interaction through the assistance of spin-orbit coupled supercurrents [44,45]. A more quantitative analysis is required to elucidate this scenario at sufficiently high excess Fe concentrations, in connection to the observed gapping of the Dirac point associated with the topological surface states [27]. The evolution of QAV with the impurity concentration is outside the scope of the current work and will be systematically examined in the future.

The TRS-broken topological vortex matter we observed may be harnessed for quantum information technology [46–51]. Because Fe(Se,Te) has shown a $Z_2$ nontrivial topological band structure and superconducting topological surface states [21,24–26], both its Abrikosov vortices and the QAVs support degenerate zero-energy excitations or Majorana zero modes, which are anyons that obey non-Abelian statistics [46–48]. Fusion and braiding of such an anionic vortex with its anti-particle are essential for fault-tolerant quantum computation [49,50]. Since free Abrikosov vortices typically have the same vorticity and thus repel each other, it will be difficult to annihilate them in order to fuse the Majorana zero modes without destroying the superconducting condensate. The perturbative nature of our local field avoids collapsing the superconducting gap and protects adiabaticity during manipulation [46,47]. Additional braiding schemes are also possible by bringing a free Abrikosov vortex around QAV-antivortex pairs using similar sSQUID manipulations.

In conclusion, we used sSQUID microscopy and directly observed a novel form of vortex with spontaneously broken TRS in Fe-based superconductor Fe(Se,Te). QAVs and its antivortices occurred stochastically at zero magnetic field due to the spin-orbit coupling between the impurity moment and vortex core-states. By applying a small local magnetic field from the nano-SQUID probe, we observed ferromagnetic-like hysteretic switching loops, following a vectorial rotation of the flux line threading the impurity spin. Our observation and manipulation of QAV may enable new possibilities for superconducting information technology in a promising material platform.

## Acknowledgement

YHW would like to acknowledge partial support by the Ministry of Science and Technology of China under Grant numbers 2017YFA0303000, NSFC Grant No. 11827805 and 12150003, and Shanghai Municipal Science and Technology Major Project Grant No. 2019SHZDZX01. YFG would like to acknowledge support by NSFC Grant No. 11874264. ZW is supported by US DOE, Basic Energy Sciences Grant No. DE-FG02-99ER45747. CZ and KL acknowledge the financial support from the NSFC Grant No. 61871134 and Shanghai Municipal Science and Technology Commission Grant No. 18JC141030. ZKL would like to acknowledge support by the National Key R&D program of China Grant No. 2017YFA0305400. All the authors are grateful for the stimulating discussions with Y. Yu, J. Shen, X. Dai, and J. Zi as well as experimental assistance by J. Zhao, D. Jiang, T. Zhang and D. L. Feng.

## Data availability

The data that support the findings of this work are available from the corresponding authors upon reasonable request.

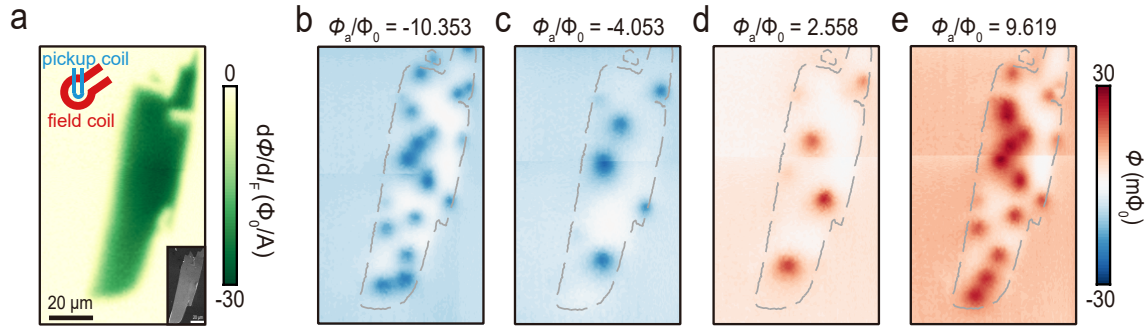

**Fig. 1 Abrikosov vortices in an Fe(Se,Te) superconductor imaged by scanning SQUID. a,** Susceptometry image of the sample. The dark green area with diamagnetic susceptibility represents the sample, while the white area represents the substrate. The orientation of our SQUID with respect to the sample is illustrated. Inset, scanning electron micrograph of the sample. **b-e,** Magnetometry images of the sample after field cooling under various applied flux $\Phi_a$ through the sample. The gray dashed lines outline the boundary of the sample determined from susceptometry. The horizontal discontinuities in the images were a result of stitching two images together. All the images were obtained at 1.5 K. In this regime of $|\Phi_a| > \Phi_0$, vortices nucleated with the same vorticity as determined by the sign of $\Phi_a$. The amount of observed vortex was clearly larger than $|\Phi_a|/\Phi_0$.

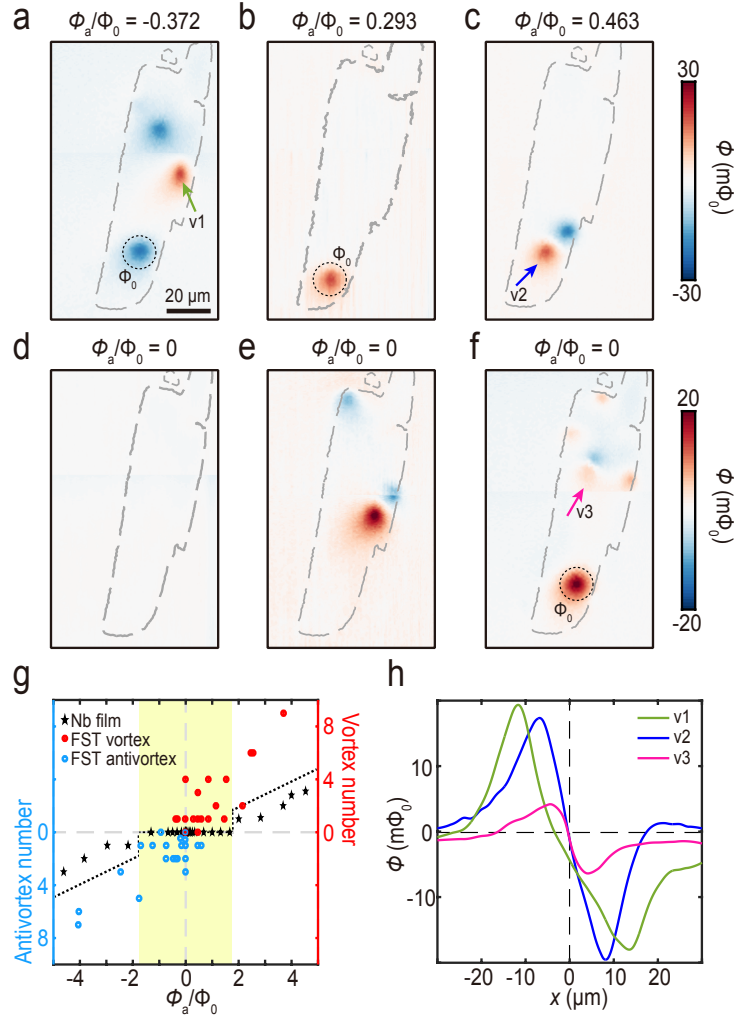

**Fig. 2 Spontaneous generation of Quantum Anomalous Vortex (QAV) in Fe(Se,Te) under zero-field cooling. a-f,** Magnetometry images of the sample after field cooling under various $|\Phi_a| < \Phi_0$, a regime no vortex was expected on a conventional superconductor. They are representatives of the 27 images obtained in this regime. In the strict zero-field cooling cases (**d-f**), the vortex-antivortex patterns appeared different each time (more in [23]). Dashed circles in **d, b** and **f** outline isolated vortices with quantized flux of $\Phi_0$. **g,** Number of vortices as a function of $\Phi_a$. The red dots and blue circles represent positive and negative vortex number respectively under each cooling cases while the black stars were a baseline from the Nb film (Fig. S5k). The yellow shaded area represents the low-field regime where no vortex was observed on Nb while spontaneous vortices appeared in Fe(Se,Te). The black dash line outside the low-field regime marks the boundary where the vortex number equals $\Phi_a/\Phi_0$. The random and spontaneous vortex formations of both vorticities suggest the nucleation of QAV and antivortex. **h,** Line-cuts of three typical vortex-antivortex patterns. The three cuts (green, cyan and magenta) are along the arrow directions v1, v2 and v3 in **a**, **c** and **f** with matching colors, respectively. The coupling between QAV-antivortex was stronger when they located closer.

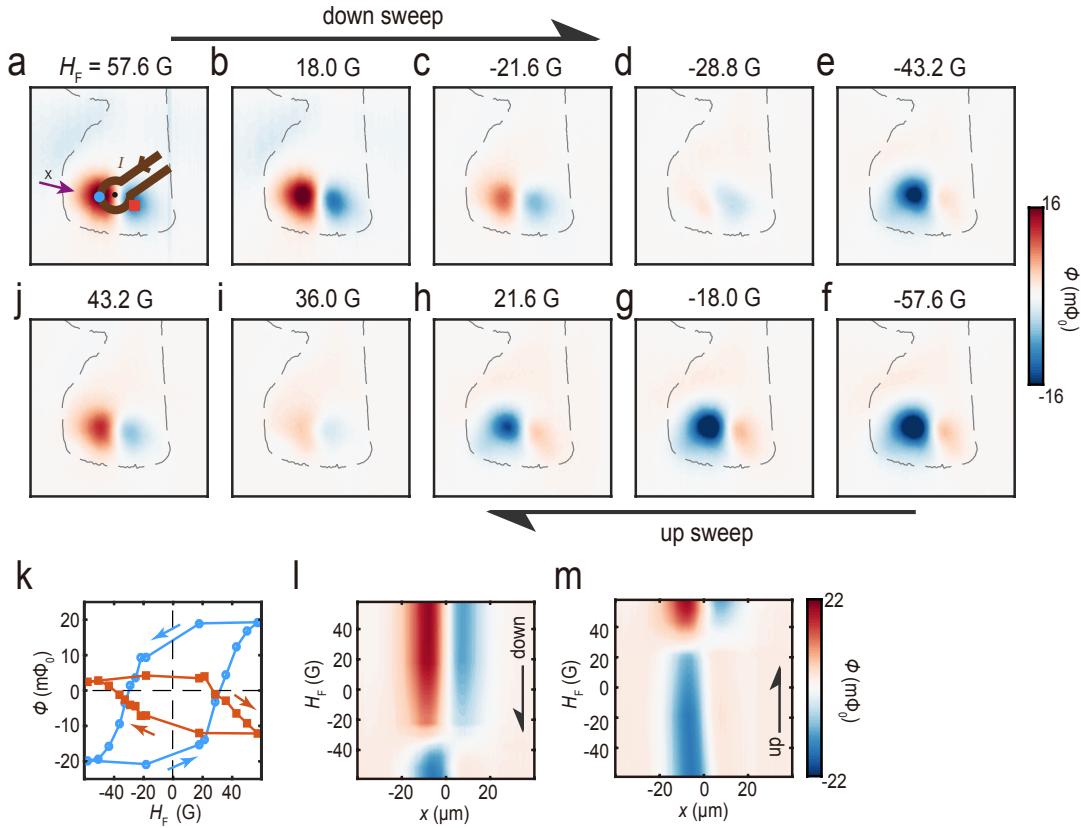

**Fig. 3 Hysteretic switching of QAV and its antivortex using a local field. a-j,** Magnetometry images of an Fe(Se,Te) sample obtained at 1.5 K and zero field. The dashed lines outline the shape of the sample as obtained from susceptometry. A local field $H_F$ was applied by passing a DC current through the field coil (**a**, brown line) on the nano-SQUID probe at the middle of the vortex-antivortex dual (**a**, black dot) and then removed before each image was taken. The history of the images is indicated by the black arrows showing field sweeping directions. **k,** Flux signals extracted from the two points (**a**, blue dot and orange square) as a function of $H_F$. The magnetic hysteresis loops exhibiting opposite winding with $H_F$ corresponds to QAV and antivortex, respectively. **l** and **m** are interpolated images from the line-cuts through the pattern in **a** (arrow direction) as a function of $H_F$ from down- and up-sweeps, respectively. The hysteresis in the switching was reminiscent of a ferromagnetic magnetization loop and suggested spontaneously broken time-reversal symmetry of the underlying order parameter.

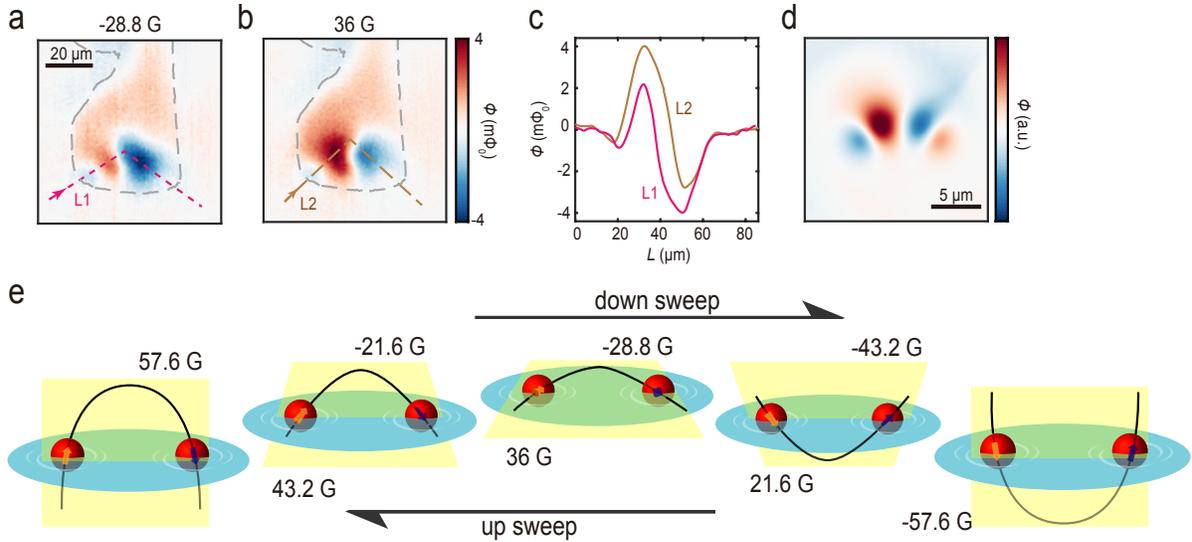

**Fig. 4 Vectorial rotation of a QAV and its antivortex. a** and **b,** Magnetometry images from Figure 3**d** and **i**, respectively, rescaled in color. **c**, line-profiles from **a** and **b** along the solid line. **d**, simulated out-of-plane magnetic field of a QAV-antivortex pair oriented in-plane (see text). **e**, A cartoon depicting a pair and its impurity (red spheres) moments (orange and purple arrows) rotating about their central axis when a local field was applied. The blue circular plate represents the sample in-plane; the black line represents the flux line which goes through the impurity moments; the yellow plane is an auxiliary plane the moments and the flux line are confined to. The fields match the corresponding fields in Figs. 3a-j. Such polar rotations of the flux line of QAV differentiated it from horizontal motion of an Abrikosov vortex by a local probe and suggested a new mechanism enabling efficient manipulation of vortex.